\documentclass[preprint,aps,showpacs,nofootinbib,tightenlines,showkeys,
    byrevtex]{revtex4}
\usepackage{graphicx}%
\usepackage{dcolumn}
\usepackage{amsmath}
\usepackage{epsfig}
\usepackage{latexsym}
\usepackage{amssymb}
\usepackage{mathrsfs}

\begin{document}
\bibliographystyle{plain}

\title{Precision calculation for nucleon capture by deuteron with
Effective Field Theory}

\author{S. Bayegan}\email{Bayegan@Khayam.ut.ac.ir}
 \affiliation{Department of Physics, University of Tehran, P.O.Box 14395-547, Tehran, Iran.}

\author{H. Sadeghi}\email{Hsadegh@Chamran.ut.ac.ir}
 \affiliation{Department of Physics, University of Tehran, P.O.Box 14395-547, Tehran, Iran.}

\begin{abstract}
\vspace{0.5cm}
 We calculate the cross section for radiative capture of neutron
 by deuteron $n+d \rightarrow 3H+\gamma$ using Effective Field
 Theory(EFT). The calculation includes N2LO order and we compare our
 results with available calculated data below E=0.2 Mev.
\end{abstract}


\maketitle

\begin{section}{Introduction}
\label{introduction}

Very low-energy radiative capture and weak capture reactions
involving few-nucleon systems have considerable astrophysical
relevance for studies of stellar structure and evolution and big-
bang nucleosynthesis. In the standard Big-Bang deuterons being to
be formed through the process $np \rightarrow d\gamma $ and then
the following reactions of primordial nucleosynthesis proceed
rapidly: $pd \rightarrow 3He\gamma$, $nd \rightarrow 3H\gamma$,...
. These reactions at the energies relevant for BBN, $0.02 < E <
0.2$ MeV, is not well-measured experimentally and there are
significant theoretical uncertainties~\cite{Burles}. Some
theoretical calculation has been down for these reactions but all
of them show very significant theoretical error.

 In this paper we show result with EFT for reaction $nd \rightarrow
3H\gamma$.  In next section we show results for two body radiative
capture process $np \rightarrow d\gamma$ and in the last section
we show our results for three body radiative capture process $nd
\rightarrow 3H\gamma$ and then we will compare these results with
EFT at N2LO with ENDF/B-VI~\cite{ENDF}.

\section{Two Body sector($np \rightarrow d\gamma$)}
An accurate estimation of the capture processes cross sections
 are essential to the BBN
prediction of deuterium and other light element abundances. For
example, $np \rightarrow d\gamma$   In the model calculation of
Smith, Kawano and Malaney (SKM) an error of $5\%$ was assigned to
the cross section for $np \rightarrow d\gamma$~\cite{SKM}. The SKM
result agrees with a slightly earlier evaluation from
ENDF/B-VI~\cite{ENDF}. The errors in this calculation are not well
documented and they could be as large as $10$-$15\%$. There is a
much earlier calculation (1967) by Fowler, Caughlan and Zimmerman
 which agrees with the ENDF/B-VI result for energies $E < 0.2$
 MeV but disagrees significantly
at higher energies.

In model independent calculation or EFT calculation for $np
\rightarrow d\gamma$ calculation by Savage and Chen (1999) show
$4\%$ theoretical uncertainly~\cite{Savage}. Precision and higher
order calculation by Rupak (2000) show $1\%$ theoretical
uncertainly for energy $\sim 1$ Mev~\cite{Rupak}.

\section{Three nucleon system in triton channel and radiative
capture of neutron by deutron }
 As discussed by Bedaque et. al.~\cite{Bedaque}, the
$^2S_{\frac{1}{2}}$ channel -- to which $^3$He and $^3$H belong --
is qualitatively different from the other three-nucleon channels.
This difference can be traced back to the effect of the exclusion
principle and the angular momentum repulsion barrier. In all the
other channels, it is either the Pauli principle or an angular
momentum barrier (or both) which forbids the three particles to
occupy the same point in space.

 The spin structure of the matrix
elements for $Nd$ radiative capture is complicate also for the low
energy interaction, as we have here 3 independent multipole
transitions (allowed by the $P-$parity and the total angular
momentum conservation): ${\cal
J}^{P}=\displaystyle\frac{1}{2}^+\to M1$ and $ {\cal
J}^{P}=\displaystyle\frac{3}{2}^+\to M1$ and $E2$, with the
following parametrization of the corresponding contributions to
the matrix element:
$$i(t^\dagger N)(\vec D\cdot\vec{e^*}\times\vec k),$$
\begin{equation}
(t^\dagger\sigma_a N)(\vec D\times [\vec{ e^*}\times\vec k])_a,
\label{eq:as7}
\end{equation}

where $N$, $t$, $\vec e$, $\vec D$ and $\vec k$ are the
2-component spinors of initial nucleon filed, final $^3He$ (or
$^3H$)field, the 3-vector polarization of the produced photon, the
3-vector polarization of deuteron and the unit vector along the
3-momentum of the photons, respectively.  The two structures in
Eq.(\ref{eq:as7}) correspond to the M1 radiation. The M1 amplitude
recived contributions from the magnetic moment of nucleon and
four-nucleon operators coupling to magnetic fields, which
described by the lagrange density involving dibaryon fields:
\begin{equation}\label{eq:M1}
  \mathcal{L}_B=\frac{e}{2M_N}N^\dagger(k_0+k_1 \tau^3){\sigma.B}
  +e\frac{L_1}{M_N\sqrt{r^{({^1s}_0)}r^{({^3S}_1)}}}{{d_t}^j}^\dagger{{d_s}}_3
  B_j+H.C.
\end{equation}
where $d_t$ is the $^3S_1$ dibaryon and $d_s$ is $^1S_0$ dibaryon.
The coefficient $L_1$ is determined from two body capture
experimental data~\cite{Rupak}.
 In Fig. (\ref{3bodyeqca})we show diagrams which can be contributed at N2LO, with expansion of the
  kernel in power of
 Q and then we can iterate the kernel by inserting into integral equation and regularize at
 N2LO we finally arrived to cross section of radiative capture of neutron
 by deuteron.
\begin{figure}[!htb]
\begin{center}
  \includegraphics[width=0.5\linewidth,angle=270,clip=true]{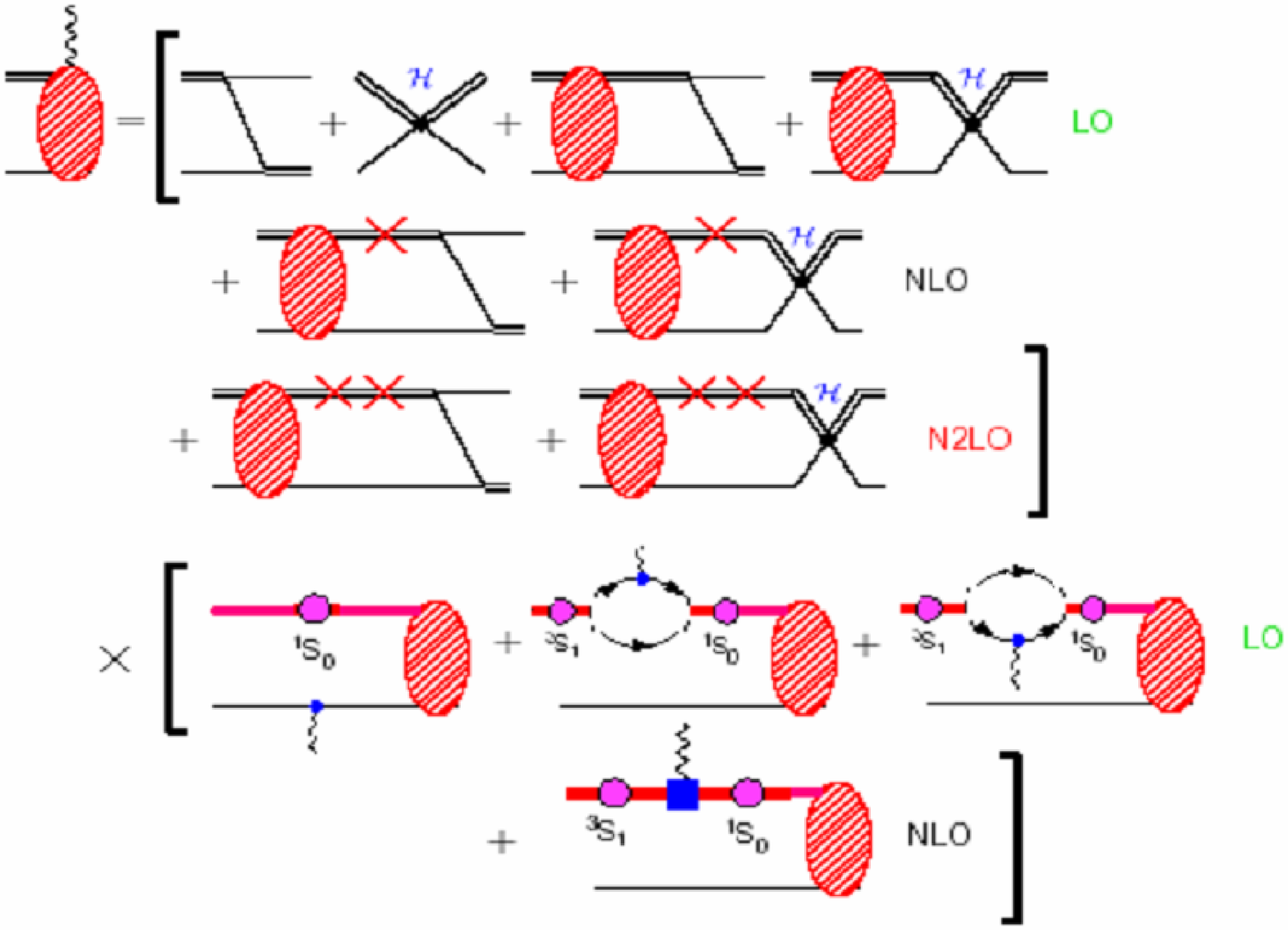}
\end{center}
\vspace*{-0pt} \caption{Diagrams contributing to the M1 amplitude
for $nd \rightarrow 3H\gamma$ at N2LO.}
\label{3bodyeqca}
\end{figure}

\section{results} We show numerical results for cross section in
Table(1). This calculation show at LO 10-40 $\%$ accuracy, at NLO
 < 10$\%$  accuracy and at N2LO  1$\%$ accuracy for energy
~50-70 Kev in comparison with Evaluated Nuclear Data
File(ENDF)~\cite{ENDF}.

\begin{table}[!htb]
\caption{Neutron radiative capture by deuteron in micro barn at
N2LO. Last column shows ENDF results for cross
section~\cite{ENDF}.} \label{tab:a} \vspace{0.25cm}
\begin{center}
\begin{tabular}{c c||c|c|c|c}
\hline &Energy  & $\sigma(\mu b)$ & $\sigma(\mu b)$ &
$\sigma(\mu b)$ & $ENDF(\mu b)$ \\
&(Kev)  & LO & NLO & N2LO &  \\
 \hline \hline

 & 40 & 1.64 & 1.31    & 1.25 & 1.27(0)\\
 & 50 & 1.72 & 1.45 & 1.38 & 1.39(0)\\
 & 60 & 1.90 & 1.58  & 1.50 & 1.50(0)\\
 & 70 & 2.02 & 1.72  & 1.61 &  1.61(0)\\
 & 80 & 2.11 & 1.82  & 1.73 & 1.72(0)\\
 & 100 & 2.42 & 2.04  & 1.93 & 1.94(0)\\
 & 140 & 2.90 & 2.42  & 2.30 & 2.22(9)\\
\hline
\end{tabular}
\end{center}
\end{table}


\begin{acknowledgments}
We thanks U. van kolck for useful comments and discussion and
special thanks to Paulo.F. bedaque and Harald W.~Grie\ss hammer
for providing useful and valuable mathematica code.
\end{acknowledgments}
\end{section}

\bibliographystyle{apsrev}

\end{document}